\documentclass[preprint2]{aastex}
\begin{document}

\title{Polaris and its Kin}

\author{David G. Turner}
\affil{Saint Mary's University, Halifax, Nova Scotia, Canada}
\email{turner@ap.smu.ca}

\begin{abstract}
A review is presented of the past 165 years of observation of the 4-day Cepheid Polaris, including the exciting results of the last 50 years, an interval that has produced three orbital solutions for the spectroscopic binary subsystem, recently resolved by HST, parameters for the optical companion, precise measurement of the star's trigonometric parallax and angular diameter, evidence for a rapid increase in its pulsation period, and observations of the dramatic decline and recent partial recovery of its light amplitude. There has been considerable discussion about the exact nature of the star, with potential resolutions summarized here. It is also noted that many of the star's characteristics are shared by a small number of other Cepheids that display rapid period increases identical to those predicted for stars in the first crossing of the instability strip, small light amplitudes, and intrinsic colors typical of variables lying near the center of the strip, where Cepheids of largest amplitude reside. While all members of the group appear to display the canonical traits of first crossers of the instability strip, Polaris has one unique peculiarity: a brief hiatus in its monotonic period increment between 1963 and 1966 during which the pulsation period underwent a dramatic decrease. Has the average brightness of the Cepheid also increased over the last millennium?
\end{abstract}

\section{Introduction}

Polaris ($\alpha$ UMi, spectral type = F7 Ib) is the nearest and brightest of $\sim40$ classical Cepheids detectable without optical aid \citep{1}, and possibly the most enigmatic. It has a pulsation period of $3^{\rm d}.97$ and the second smallest light amplitude \citep{2}. Its light variability was suspected by \citet{3} from measures he made using a Steinheil eyepiece photometer that were corrected for atmospheric extinction and had a precision of about $\pm0^{\rm m}.05$. The accuracy of measures made with the photometer is somewhat uncertain \citep{4}, and in any case awaited the full development of the magnitude scale. Variability was also suspected in visual comparisons of the star with $\beta$ UMi made by Schmidt between 1843 and 1856 \citep[1144 estimates,][]{5} and Pannekoek between 1890 and 1894 \citep[510 estimates,][]{6}, but confirmation of the star's light variations would await photographic observations by \citet{7}, after the star's 4-day cycle of radial velocity variability had been announced previously \citep{8}. The original light amplitude was very small, only $\sim10\%$, which is why it remained unnoticed for so long. In the {\it General Catalogue of Variable Stars} it was listed as a Type II Cepheid as recently as 1970 \citep{9}, although its metallicity \citep{10} and distance (see below) confirm it to be a classical Cepheid.

Polaris is a triple system \citep{11,12} consisting of an unresolved F6 V dwarf, Polaris Ab, in a 30-year spectroscopic binary orbit and a F3 V dwarf (Polaris B) \citep{13,12}, $18^{\prime\prime}$ distant, as a visual physical companion. Polaris has a precise {\it Hipparcos} parallax of $\pi_{\rm abs} = 7.72 \pm0.12$ mas \citep{14} that yields a distance of $130 \pm2$ pc. The corresponding luminosity, $\langle M_V\rangle = -3.62 \pm0.05$, implies overtone (OT) pulsation, consistent with the Cepheid's sinusoidal light curve. But there is no indication in Fourier analyses of photometric or radial velocity measures of a signal at the putative fundamental mode (FM) period of $5^{\rm d}.7$ \citep{15,16,17}, and a sinusoidal light curve may not always imply overtone pulsation \citep{18}. Classification through Fourier parameterization is not necessarily reliable for a light curve of extremely small amplitude consistent with a pure sine wave, and a preliminary investigation of nine other sinusoidal Cepheids associated with open clusters \citep{19} indicates that only four are likely OT pulsators. V1726 Cyg in the cluster Platais 1 is a good example of what might be the case for Polaris: a low-amplitude, sinusoidal Cepheid pulsating in the FM with $\langle M_V\rangle = -3.08 \pm0.07$ \citep{20}. What is the situation for Polaris?

\section{Background Information}

\begin{figure}[ht]
\includegraphics[width=0.2\textwidth]{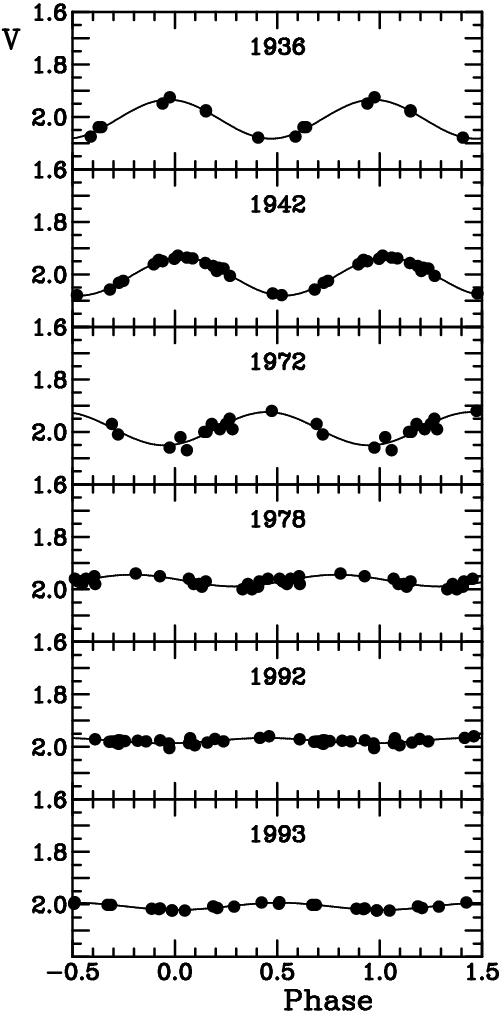}\includegraphics[width=0.25\textwidth]{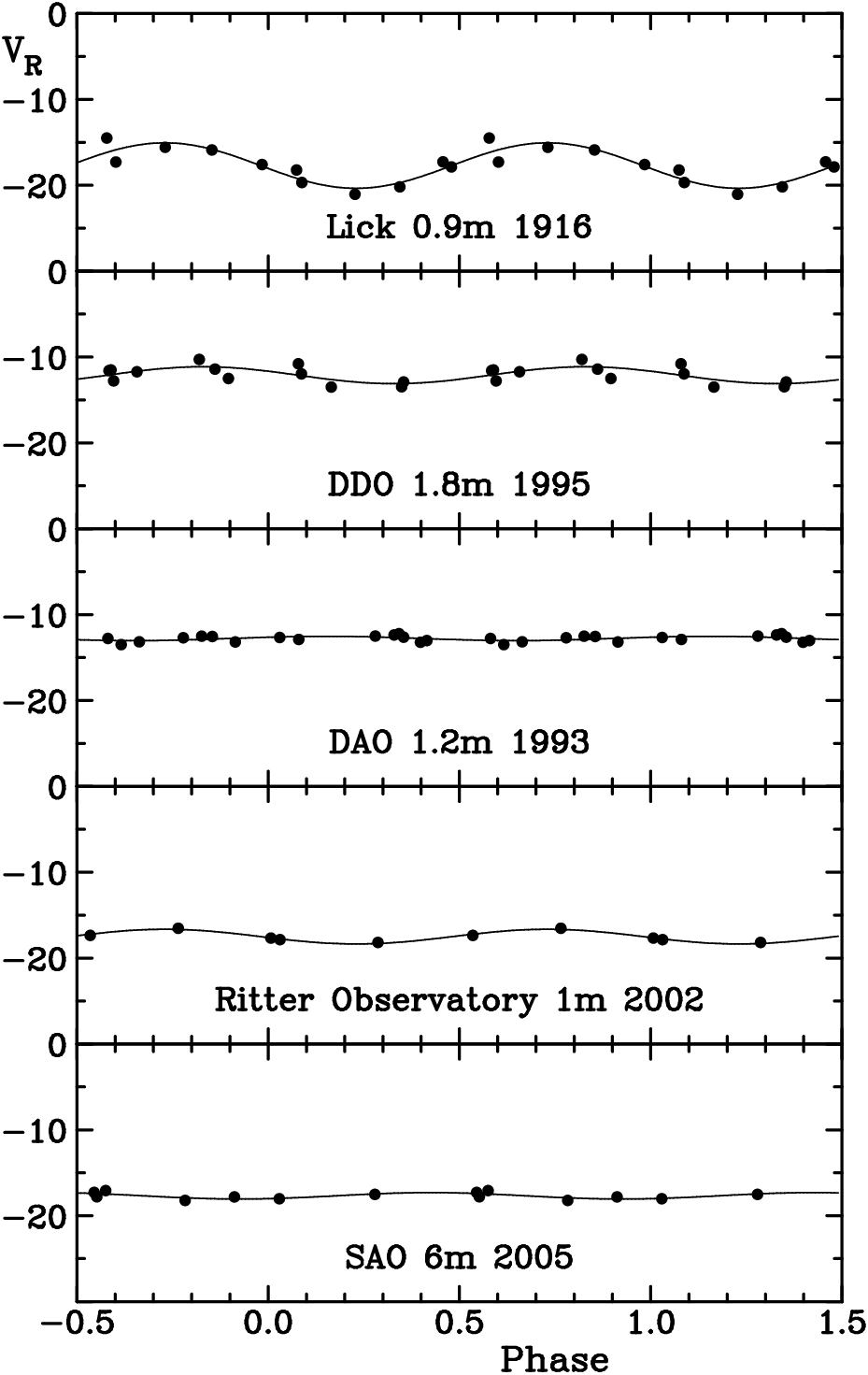}
\caption{\label{fig1} \small{Photometric light curves for Polaris (left) from: \citet{25}, \citet{26}, \citet{27}, \citet{28}, and \citet{29}; phased radial velocity variations (right) from \citet{24}, \citet{11}, and \citet{30}, with source telescopes indicated.}}
\end{figure}

Polaris illuminates diffuse dust nebulosity \citep{21} as well as a reflection nebula $\sim1^{\circ}$ south \citep{22}, presumably the associated dust clouds accounting for the Cepheid's small reddening of $E_{B-V} = 0.02\pm0.01$ \citep{23,10,15}. The star has been the object of numerous ground-based investigations of its variability \citep{1}, ranging from visual, photographic, and photoelectric brightness estimates to radial velocity measures \citep{24,11,1}. A selection of some of the best photometric observations is shown in Fig.~\ref{fig1}, along with a variety of different radial velocity measures. Not illustrated are the many low-quality observations that have been made over the past century \citep{1}.

Recent years have witnessed great improvements in the quality of both the photometric and radial velocity monitoring of Polaris, including observations from space \citep{16,17} and high resolution spectroscopy \citep{11,31,30,32}.
High quality radial velocity measurements are probably the optimum method for updating information on the period and amplitude changes of Polaris from ground-based observatories, although it is from exactly such data that evidence has been presented for an additional periodicity in the variability of Polaris, other than that arising from its pulsational and orbital motions. Is the 119-day periodicity found by \citet{32} evidence for a fourth star in the system, or is it attributable to seasonal effects on ground-based equipment? A similar feature was detected by \citet{33} and \citet{31} in their radial velocity studies of Polaris, but with a shorter period near 40 days.

\section{Period and Amplitude Changes}
\begin{figure}[h]
\includegraphics[width=0.4\textwidth]{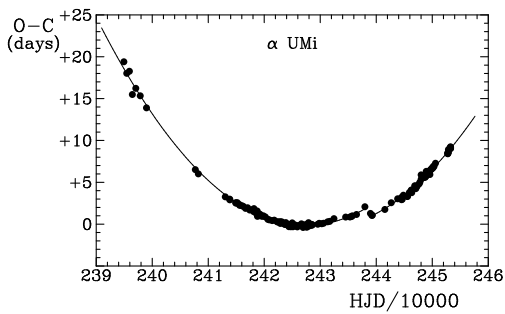}
\caption{\label{fig2} \small{O--C variations of Polaris from photometric and radial velocity data, without correction for orbital motion.}}
\end{figure}

A recent study of the period changes in Polaris \citep{1} updated and expanded upon more limited investigations by \citet{34} and others \citep{35,29,36}. The \citet{1} study included light travel time corrections for the orbital motion of Polaris about its F6 V companion \citep{11}, although it was noted in the analysis that such corrections did not appear to remove the effects entirely. The original analysis was redone for the present study with the inclusion of recent observations, but excluding the original light travel time corrections, in order to determine if they could be established independently. The results are presented in Fig.~\ref{fig2}.

The long-term trend in the pulsation period of Polaris from O--C analysis is that it is constantly increasing, except for an unusual ``glitch'' circa 1965, during a gap in the observational coverage between 1963 and 1966. Prior to the glitch, the rate of period increase is calculated to have been $4.46 \pm0.03$ s yr$^{-1}$. Following the glitch it is calculated to be $4.19 \pm0.13$ s yr$^{-1}$. Both values are consistent with results from stellar evolutionary model calculations for a star in the first crossing of the instability strip \citep{37}, provided that Polaris is a FM pulsator. If the Cepheid is in another crossing of the instability strip, say a third or fifth crossing, the observed rates of period increase are several times larger than predicted from model predictions \citep{37}. The shift in the O--C data during the glitch corresponds to an abrupt decrease in pulsation period, which, according to simple calculations tied to the period-density relation, could be accommodated through the sudden acquisition by the Cepheid of about seven Jovian masses of matter \citep{1}. The true explanation remains a mystery, although it is interesting to note that the assimilation of any pre-existing planetary companions to Polaris is most likely to occur during the first crossing of the instability strip, when the star is growing to red supergiant dimensions for the first time.

\begin{figure}[h]
\includegraphics[width=0.4\textwidth]{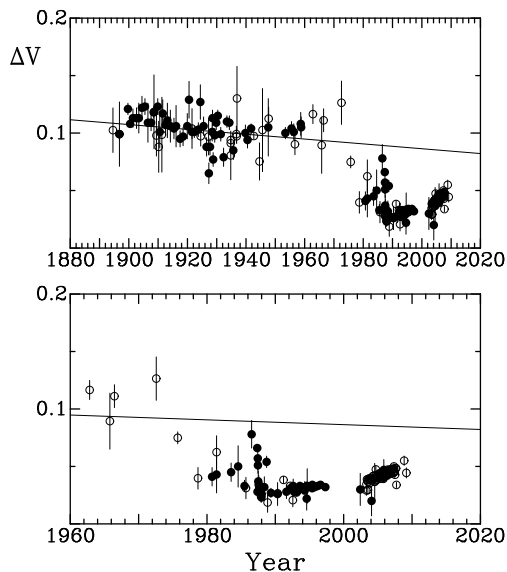}
\caption{\label{fig3} \small{{\it V}-band amplitude variations of Polaris from photometric (open circles) and radial velocity (filled circles) data, with uncertainties indicated. The straight line is a weighted fit to the pre-1966 data. The lower portion expands the data for the past 50 years.}}
\end{figure}

The amplitude variations in {\it V} of Polaris are plotted in Fig.~\ref{fig3}, which includes radial velocity amplitudes scaled by the usual factor of 50 km s$^{-1}$ magnitude$^{-1}$.  The light amplitude prior to the 1965 glitch appears to have been undergoing a slow but steady decline, with the much more rapid decline noted by others \citep{34,33,29,36} beginning roughly a decade later, reaching minimum at $\Delta V = 0^{\rm m}.025$ circa 1988. It is presently increasing, but may decay completely by about 2400 if the pre-glitch decline continues.

A longer-term overall brightening of $\sim 1^{\rm m}.7$ (from 3.6 to 1.9 in {\it V} from $\sim 100$ CE to the present) has been suggested \citep{38,39}, based upon an analysis of archival and recent brightness estimates. Such a gradual brightening over the last millennium would make Polaris unique among Cepheids, but appears to conflict with two sets of observations: those of \citet{5} that imply $\langle V \rangle \simeq 2.0$ provided there is no color term in his comparison of Polaris with the much redder $\beta$ UMi \citep{1}, and those of \citet{40} over the past seven years that imply $\langle V \rangle = 1.991\pm0.004$ s.d., a value which includes slight contamination from the light of Polaris B amounting to less than $0^{\rm m}.006$ ({\it i.e.}, $\langle V \rangle \simeq 1.997$ for the Cepheid alone). In other words, the mean brightness of Polaris appears to have been constant at $V \simeq 2.00$ at least over the past century and a half.

\section{Basic Parameters}

A model atmosphere analysis of Polaris by \citet{10} finds an abundance pattern with [C/H] = --0.17, [N/H] = +0.42, [O/H] = --0.00, and [Na/H] = +0.09, with the nitrogen and sodium enhancement and carbon (and oxygen) depletion being a strong signature of CNO-processed elements. \citet{10} argue that Polaris is a post red supergiant dredge-up star because of such a signature. Yet the presence of CNO-processed elements in Cepheid atmospheres appears to be unrelated to red supergiant dredge-up, which does not take place in some evolutionary models, the contamination occurring instead in late main-sequence phases \citep{41,42}. The most likely mechanism is meridional mixing in rapidly-rotating B-type stars. As noted earlier, the rapid rate of redward evolution indicated by its rate of period increase suggests that Polaris is in the first crossing of the instability strip \citep{37}. The rate is almost two orders of magnitude larger than predicted by stellar evolutionary models for a third crossing, although a fifth crossing might be an outside possibility.

The angular diameter of Polaris has been measured with the Naval Prototype Optical Interferometer \citep{43}, yielding best estimates of $\theta_{\rm UD} = 3.14 \pm0.02$ mas and $\theta_{\rm LD} = 3.28 \pm0.02$ mas. Cepheids are known to obey a tight period-radius (PR) relation established from Baade-Wesselink and surface brightness methods \citep{44,45,46}, with $\langle R \rangle \propto P^{\frac{3}{4}}$. Cepheids also appear to obey a period-mass (PM) relation established from cluster membership \citep{47}, with $M \propto P^{\frac{1}{2}}$. For the PR relation of \citet{46}, the implied distance to Polaris from its angular diameter is $93 \pm2$ pc ($\theta_{\rm LD}$) or $97 \pm2$ pc ($\theta_{\rm UD}$) for FM pulsation, and $122 \pm3$ pc ($\theta_{\rm LD}$) or $128 \pm3$ pc ($\theta_{\rm UD}$) for OT pulsation. For comparison, the {\it Hipparcos} parallax \citep{14} corresponds to a distance of $130 \pm2$ pc, while association with Polaris B implies distances of $101\pm3$ pc \citep{15} or $109.5$ pc \citep{48}. The limb-darkened angular diameter of Polaris does not appear to produce a distance estimate via this prescription that is entirely consistent with either of the values established by its cluster or trigonometric parallaxes, which is puzzling.

Polaris has a measured surface gravity of $\log g = 2.0 \pm0.3$ \citep{10} from LTE model atmospheres. The corresponding mass estimates for Polaris using the relations described above are given in Table~\ref{tab1}. All of the estimates appear reasonable, and imply a mass for the Cepheid lying somewhere between 4 and 7 $M_{\odot}$.

\begin{deluxetable}{ccc}
\tabletypesize{\small}
\tablecaption{Mass estimates for Polaris} 
\label{tab1}
\tablewidth{0pt}
\tablehead{\colhead{$M/M_{\odot}$}  &\colhead{Pulsation Mode} &\colhead{Technique} }
\startdata
$3.9 \pm2.9$ &FM &$\log g$ and PR relation \citep{46} \\
$6.6 \pm4.9$ &OT &$\log g$ and PR relation \citep{46} \\
$5.1 \pm0.5$ &FM &Cluster PM relation \citep{47} \\
$6.1 \pm0.6$ &OT &Cluster PM relation \citep{47} \\
$4.5 \pm1.8$ &OT? &{\it Hipparcos} $\pi$ and orbit \citep{12} \\
\enddata
\end{deluxetable}

\section{The Polaris Cluster}

\begin{figure}[h]
\begin{center}
\includegraphics[width=0.3\textwidth]{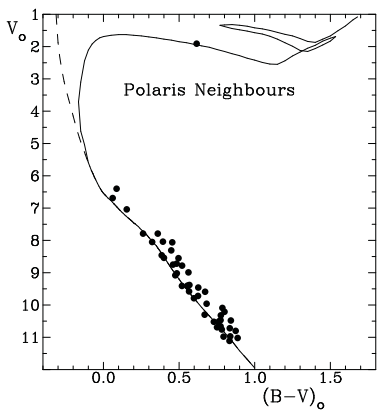}
\caption{\label{fig4} \small{{\it BV} photometry from {\it Hipparcos} for stars within $3^{\circ}$ of Polaris. The ZAMS is plotted for $d = 99$ pc and $E_{B-V} = 0.02$, along with an isochrone for $t = 8 \times 10^7$ yrs.}}
\end{center}
\end{figure}

The field surrounding Polaris appears to coincide with the remains of a poorly-populated star cluster. For example, the radial velocities of Polaris and its brightest companions are nearly identical: $-16$ km s$^{-1}$ \citep[Polaris,][]{24,11,30}, $-15$ km s$^{-1}$ \citep[Polaris B,][]{11,48}, and $-11$ km s$^{-1}$ \citep[HD 5914,][]{49}. A color-magnitude diagram for possible cluster stars from {\it Hipparcos} photometry is shown in Fig.~\ref{fig4}. Most of the stars, other than Polaris, have parallaxes corresponding to distances of $\sim 100$ pc. The age isochrone for $t = 8 \times 10^7$ years shown in Fig.~\ref{fig4} was chosen to correspond to Polaris as a first-crossing Cepheid.

\begin{figure}[h]
\begin{center}
\includegraphics[width=0.3\textwidth]{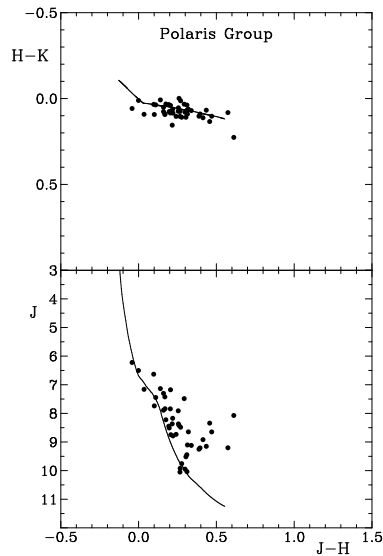}
\caption{\label{fig5} \small{{\it JHK} photometry \citep{50} for stars in Fig.~\ref{fig4}. The ZAMS in this case is plotted for $d = 106$ pc for $E_{B-V} = 0.02$.}}
\end{center}
\end{figure}

Similar results are obtained using 2MASS photometry \citep{50}, as shown in Fig.~\ref{fig5}, although in this case the implied distance from the {\it JHK} color-color and color-magnitude diagrams is $106 \pm7$ pc for $E_{B-V} = 0.02$.

Polaris is also argued to be a member of the Pleiades moving group \citep{38}, for which the main-sequence turnoff mass is 5.9 $M_{\odot}$. We have been unable to confirm a similarity in {\it U}, {\it V}, and {\it W} motions between Polaris and Pleiades members from an independent analysis, so it is unclear if such a connection exists, despite the close agreement in implied evolutionary ages for Polaris and Pleiades members.

\section{Polaris Kin?}
\citet{51} argues that Polaris and a few other unusual Cepheids (e.g., V473 Lyr) may constitute a special class of ``Blazhko Cepheids,'' on the basis of their unusual period and amplitude changes, akin to the Blazhko Effect in RR Lyrae variables. In this mechanism the growth and decay of a strong magnetic field modulates surface convection, thereby modulating pulsation amplitude and period. For stars like Polaris the pulsation period should increase as amplitude declines (stronger convection), and decrease as amplitude rises (weaker convection). In Polaris the period changes are in the exact opposite sense, so the glitch and subsequent decline in light amplitude for the Cepheid over the past thirty years remain unexplained by such a mechanism. In fact, the dominant period changes observed in the large majority of Cepheids are readily explained by the evolutionary changes in radius that occur as massive stars evolve through the Cepheid instability strip \citep{37}.

\begin{figure}[h]
\includegraphics[width=0.4\textwidth]{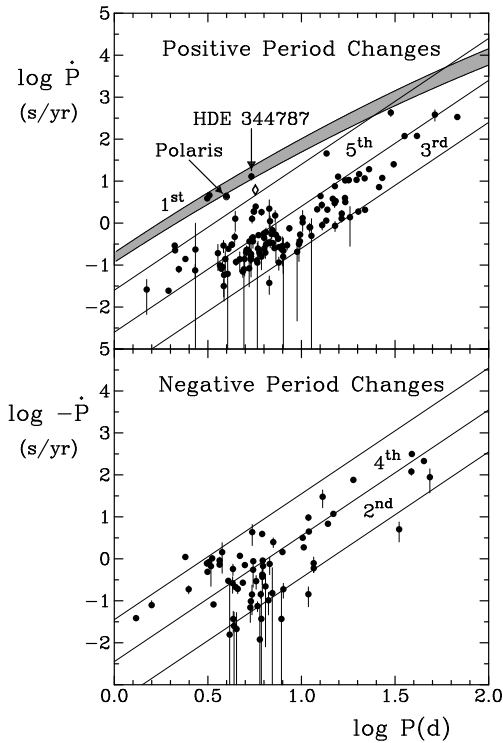}
\caption{\label{fig6} \small{Observed rates of period change for Cepheids \citep{37}, displaying the likely association with instability crossing mode: third, fifth (top), second, fourth (bottom). The shaded area corresponds to evolutionary model predictions for a first crossing. Data points corresponding to Polaris and HDE 344787 are identified, that for the most recent rate of period increase for Polaris as an open circle (below the filled circle datum). Open diamonds represent where the corresponding period changes of Polaris would fall if it is an OT pulsator.}}
\end{figure}

It is possible to map the rate of period change diagram ($\dot P - \log P$) from stellar evolutionary models for individual crossings of the Cepheid instability strip, as in Fig.~\ref{fig6}. The rapid first crossing occurs when the stars are undergoing rapid core contraction supported only by a surrounding, thin, hydrogen-burning shell. The crossing occurs at a rate almost two orders of magnitude faster than the slower second and third crossings, when the stars are undergoing core helium-burning, so a random sample of Cepheids should contain only a few objects in the first crossing, perhaps only a few percent at most. Stellar evolutionary models predict specific rates of period increase for first crossings, identified by the gray area in Fig.~\ref{fig6}. Three Cepheids fall within that region: DX Gem ($P = 3^{\rm d}.14$), BY Cas ($P = 3^{\rm d}.22$), and the double-mode pulsator HDE 344787 ($P1/P0 =  3^{\rm d}.80/5^{\rm d}.40$). DX Gem and BY Cas are suspected FM pulsators, the former as a possible outlying member of an anonymous open cluster. BY Cas lies near the cluster NGC 663, but seems unlikely to be a member.

\begin{figure}[h]
\begin{center}
\includegraphics[width=0.3\textwidth]{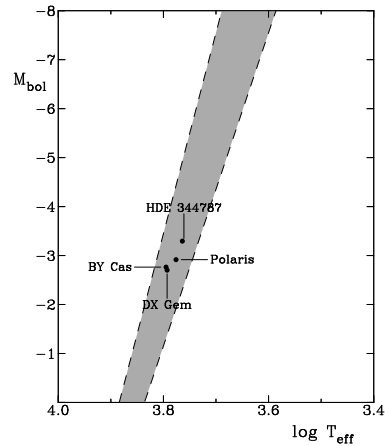}
\caption{\label{fig7} \small{The location of the putative first crossing Cepheids in the instability strip (shaded region) as FM pulsators. The strip boundaries are observationally based \citep{15}.}}
\end{center}
\end{figure}

The rates of period change for DX Gem, BY Cas, and HDE 344787 are consistent with expectations for a first crossing of the instability strip. Is that also true for Polaris? All four Cepheids have common characteristics: rapid period increases, small light amplitudes ($\Delta B = 0.49$ for DX Gem, $\Delta B = 0.53$ for BY Cas, $\Delta B = 0.06$ for Polaris, and $\Delta B \simeq 0.02$ for HDE 344787), and sinusoidal light curves. The putative first crossers also lie near the middle of the instability strip, according to their derived reddenings (Fig.~\ref{fig7}). That is consistent with model predictions from \citet{52} for stars lying towards the cool edge of the instability strip for first crossers. Alternatively, if the location of Polaris in Fig.~\ref{fig7} is adjusted upwards by, say, $0^{\rm m}.6$ to account for possible overtone pulsation \cite[see][]{1}, an anomaly arises: the star would still lie redward of the hot edge of the instability strip, but in a region where the largest amplitude pulsators are located \citep{53,37}. There is very little leeway possible in such an interpretation, given the well-established spectral type and interstellar reddening of Polaris.

The interpretation of DX Gem, BY Cas, and Polaris as putative first crossers is not without problems. DX Gem has an inferred rate of period increase of $\dot P = 3^{\rm s}.87 \pm 0^{\rm s}.19$ yr$^{-1}$, consistent with a first crossing, but with strange O--C wanderings \citep{54} and possible amplitude changes that may display a slow overall decrease with time. Its characteristics may match those proposed by Stothers for ``Blazhko Cepheids'' \citep{51}. BY Cas has an inferred rate of period increase of $\dot P = 4^{\rm s}.60 \pm 0^{\rm s}.13$ yr$^{-1}$, also consistent with a first crossing. But its O--C variations are in need of confirmation, since both recent and archival observations display peculiarities suggesting that the rate of period change undergoes modulations with time. Possible temporal changes in its light amplitude have never been investigated. In the case of Polaris the observed rate of period increase is only marginally consistent with expectations for a first crossing, unless the rates predicted from stellar evolutionary models are systematically too large.

\begin{figure}[h]
\includegraphics[width=0.4\textwidth]{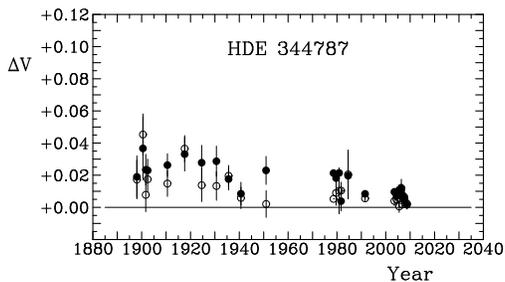}
\caption{\label{fig8} \small{The declining light amplitude of HDE 344787 for FM pulsation (open circles) and OT pulsation (filled circles) \citep{2}.}}
\end{figure}

HDE 344787 is perhaps the best case for a first-crossing Cepheid \citep{2}. It is a double-mode pulsator with an observed rate of period increase of $\dot P = 12^{\rm s}.96 \pm 2^{\rm s}.41$ yr$^{-1}$, but also with a declining light amplitude (Fig.~\ref{fig8}) that may lead to a complete cessation of pulsation within a few years' time \citep{40}.

The case for OT pulsation in Polaris may also be addressed observationally by another method. There is a well-known phase lag, $\Delta \phi_1$, between maximum brightness in Cepheid light curves and the corresponding phase of velocity minimum. \citet{55} have found that the observed period dependence of the values of $\Delta \phi_1$ for OT pulsators appears to differ from that of FM pulsators, thereby enabling one to discriminate between FM and OT pulsation in Cepheids. The general trends appear to be confirmed in models by \citet{56}. According to this criterion, the value of $\Delta \phi_1 = -0.45 \pm0.06$ obtained for Polaris by \citet{57} implies overtone pulsation. But the rapid period changes in Polaris complicate the determination of $\Delta \phi_1$ for the Cepheid, and possibly others as well. The study by \citet{1} accounted for such changes directly by carefully matching photometric and radial velocity O--C estimates closely adjacent temporally, and yielded a value of $\Delta \phi_1 = -0.10 \pm0.01$ ($-0.383 \pm0.027$ days), consistent with the FM sequence of \citet{55}, as well as that of \citet{58}. By this criterion, Polaris would be considered as a fundamental mode pulsator.

As is ever the case with Polaris, the situation is somewhat more complicated. Radial velocities also exist for the newly-discovered small amplitude Cepheid HDE 344787 \citep{2}, which appears to display different phase lags for the FM and OT modes, although the value obtained for FM pulsation is uncertain because the FM signal is much weaker than the OT mode signal. For the FM pulsation in HDE 344787 ($P = 5^{\rm d}.40$), $\Delta \phi_1 = -0.32$, whereas for the OT pulsation ($P = 3^{\rm d}.80$), $\Delta \phi_1 = -0.09$. The value of $\Delta \phi_1$ for the fundamental mode is roughly consistent with the trend found by \citet{55}, but not with that of \citet{58}, whereas the value of $\Delta \phi_1$ for the overtone mode disagrees markedly with the trends of both \citet{55} and \citet{56}. If the corresponding situation for Polaris is decided in empirical fashion, {\i.e.}, by comparison with HDE 344787, then the similarity of the value of $\Delta \phi_1$ for OT pulsation in HDE 344787 with the value of $\Delta \phi_1 = -0.10$ for Polaris, which has a similar pulsation period, suggests that Polaris is likely to be an OT pulsator.

Or is it necessarily true that double-mode Cepheids actually differ from single-mode pulsators in their $\Delta \phi_1$ properties? It may be that there is no distinction in $\Delta \phi_1$ between FM and OT pulsation, but further careful study of the observational sample is needed to confirm that possibility. The contradictions arising from the arguments presented here clearly argue for the need of improved observational data on $\Delta \phi_1$ for Cepheids of known pulsation mode, in particular data that implicitly account for the rapid period changes in many of the variables, a factor that directly affects the determination of reliable phase shifts for them.

\section{The Polaris Multiple System}
The recent resolution of the Polaris A system by \citet{12} using the {\it Hubble Space Telescope} (HST) has made it possible to study the three previously-recognized components of the system directly. The two measured HST positions for Polaris Ab relative to the Cepheid are consistent with the astrometric orbital solution proposed by \citet{59}, in which a retrograde orbit with an orbital inclination of $i = 130^{\circ}.2$ was obtained. An orbital solution by \citet{12} combined HST observations with astrometric data to yield the parameters in Table~\ref{tab2}. Matching the results to stellar evolutionary models presented some inconsistencies with the inferred mass of the Cepheid \citep{12}, possibly because it was assumed that Polaris is in the third crossing of the instability strip, which, as noted earlier, conflicts with the observed rapid rate of period change.

\begin{deluxetable}{lc}
\tabletypesize{\small}
\tablecaption{Orbital Parameters for Polaris A \citep{12}}
\label{tab2}
\tablewidth{0pt}
\tablehead{\colhead{Parameter}  &\colhead{Value}}
\startdata
Orbital inclination, $i$  &$128^{\circ} \pm21^{\circ}$ \\
Ascending node, $\Omega$  &$19^{\circ} \pm15^{\circ}$ \\
Semi-major axis, $a$  &$0^{\prime\prime}.133 \pm0^{\prime\prime}.015$ \\
Total system mass, $M_{\rm tot}$  &$5.8^{+2.2}_{-1.3}$ $M_{\odot}$ \\
Primary mass, $M_{\rm Aa}$  &$4.5^{+2.2}_{-1.4}$ $M_{\odot}$ \\
Secondary mass, $M_{\rm Ab}$  &$1.26^{+0.14}_{-0.07}$ $M_{\odot}$ \\
\enddata
\end{deluxetable}

\begin{figure}[h]
\includegraphics[width=0.45\textwidth]{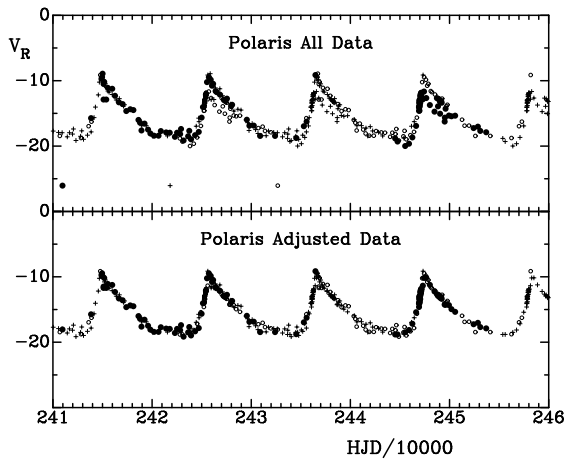}
\caption{\label{fig9} \small{Seasonal mean radial velocities for Polaris from 1888 to 2005. Filled circles represent the actual data, while open circles and crosses denote the same data folded forwards and backwards in time by one or two orbital cycles, respectively. The upper diagram illustrates the confusion introduced by literature velocities that do not appear to match the velocity system of Lick Observatory exactly. The lower diagram illustrates the data following the adoption of corrections to adjust the velocities to the Lick system.}}
\end{figure}

\begin{figure}[h]
\includegraphics[width=0.4\textwidth]{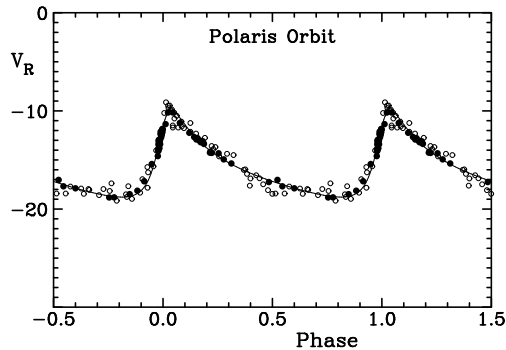}
\caption{\label{fig10} \small{The new orbital solution for Polaris A, shown as the solid curve in the diagram. Filled circles denote post-1970 data, open circles pre-1970 data.}}
\end{figure}

It is unclear how the results depend upon the orbital solution for the spectroscopic subsystem. The \citet{12} analysis of Polaris A adopted an earlier solution for the spectroscopic binary from \citet{11}, but a new solution for the system has been obtained using more recent radial velocity measures \citep{30}. The new orbital solution depends upon small zero-point adjustments to existing radial velocities, most notably the observations by \citet{11} and \citet{60}, as illustrated in the upper portion of Fig.~\ref{fig9}. Correction for such zero-point offsets reduces the scatter in the velocity observations from one cycle to another (Fig.~\ref{fig9}, lower), but leaves residual scatter that is frequently larger than expected from the uncertainties in the measurements. The resulting phased best data set is shown in Fig.~\ref{fig10} and is summarized in Table~\ref{tab3}. Three separate solutions were made by the Lehmann-Filh\'{e}s technique: one tied to recent observations, a second to the full data set (which gives a better match to the velocity amplitude {\it K}), and a third restricted to data that give the best overall visual match to the main velocity trend. An independent solution was made with the same data using a Bayesian Markov Chain Monte Carlo analysis \citep{61}. The results, shown in the last column of Table~\ref{tab3}, confirm the alternate solution based on the same data set \citep{30}. 

\begin{figure}[h]
\includegraphics[width=0.4\textwidth]{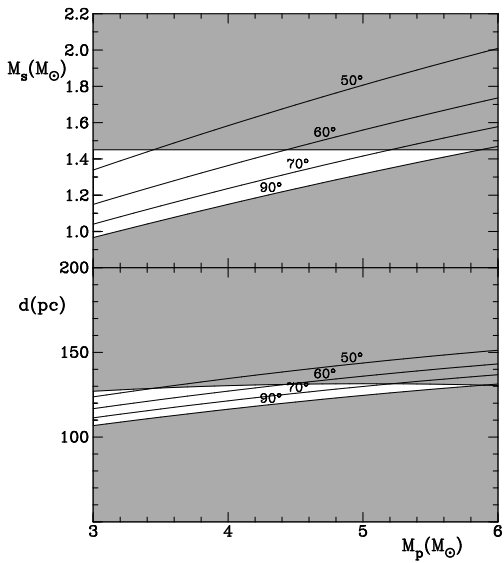}
\caption{\label{fig11} \small{Permitted parameters for the Polaris system (the unshaded region) according to the orbital solution and HST detection of the companion. The sloped curves denote, from top to bottom in each plot, orbital inclinations of $50^{\circ}$, $60^{\circ}$, $70^{\circ}$, and $90^{\circ}$, respectively. Shaded regions denote implausible orbital solutions dictated by the observed faintness of the companion (upper cutoff) and the limiting orbital inclination of $i = 90^{\circ}$ (lower cutoff).}}
\end{figure}

The new solution implies a less elliptical and larger orbit than was found in previous studies, but recent observations display kinks in the radial velocity curve for the orbit near velocity maximum that argue for the need of additional velocity coverage. It is possible to use the new orbital solution with the recent visual resolution of the companion \citep{12} to constrain the distance to the system. For this analysis the mass of Polaris is assumed to lie between 3 and 6 $M_{\odot}$, while the mass of the F6 V companion is assumed to be less than 1.45 $M_{\odot}$ \citep{12}. Such constraints produce the results shown in Fig.~\ref{fig11}.

\begin{deluxetable}{ccccc}
\tabletypesize{\scriptsize}
\tablecaption{Orbital Solutions for Polaris A}
\label{tab3}
\tablewidth{0pt}
\tablehead{\colhead{Parameter} &\colhead{\citet{24}} &\colhead{\citet{11}} &\colhead{\citet{30}} &\colhead{\citet{61}} \\ & & & & BMCMC Analysis}
\startdata
Period, {\it P} (years) &$30.46 \pm0.10$ &$29.59 \pm0.02$ &$29.71 \pm0.09$ &$29.80 \pm0.05$ \\
(days) & & &$10852 \pm33$ &$10885 \pm17$ \\
Orbital eccentricity, $e$ &$0.639 \pm0.012$ &$0.608 \pm0.005$ &$0.543 \pm0.010$ &$0.52 \pm0.01$ \\
Epoch of periastron &$1928.48 \pm0.08$ &$1928.48 \pm0.08$ &$1928.57 \pm0.06$ \\
JD & & &$2425453 \pm22$ &$2425430 \pm44$ \\
Longitude of node, $\omega$ ($^{\circ}$) &$307.24 \pm1.82$ &$303.01 \pm0.75$ &$309.6 \pm0.7$ &$301 \pm2$ \\
Semi-amplitude, $K$ (km s$^{-1}$) &$4.09 \pm0.10$ &$3.72 \pm0.03$ &$4.41 \pm0.07$ &$4.23 \pm0.07$ \\
Systemic velocity, $\gamma$ (km s$^{-1}$) &$-16.41$ &$-16.42 \pm0.03$ &$-15.90 \pm0.06$ &$-15.87 \pm0.05$ \\
Projected semi-major axis, $a_1 \sin i$ (A.U.) &$3.22 \pm0.10$ &$2.90 \pm0.03$ &$3.69 \pm0.09$ &$3.62 \pm0.06$ \\
\enddata
\end{deluxetable}

The results of the analysis can be summarized as follows. The allowable range of parameters yields $\langle d \rangle = 121 \pm10$ pc and $\langle M_V \rangle = -3.49 \pm0.18$ for Polaris, consistent with the {\it Hipparcos} parallax \citep{14} and overtone pulsation. However, the most likely orbital inclination is $i \ge 60^{\circ}$, which conflicts with the astrometric estimate of $i = 50^{\circ}$ \citep{59} unless $M_{\rm Polaris} \le 3.5 M_{\odot}$. The permitted solutions are inconsistent with the inferred distance of the visual companion (Polaris B) \citep{13,15,48}, the observed rapid rate of evolutionary period change \citep{1}, the lack of any signal tied to putative FM pulsation around $5^{\rm d}.7$ in the light and radial velocity observations \citep{15,16,17,32}, and possible membership in the surrounding anonymous cluster \citep{15}. The preferred solutions cluster near $i = 90^{\circ}$, although an edge-on orbit is inconsistent with the HST observations \citep{12}. In other words, no solution for the system parameters is ideal.

\begin{figure}[h]
\includegraphics[width=0.45\textwidth]{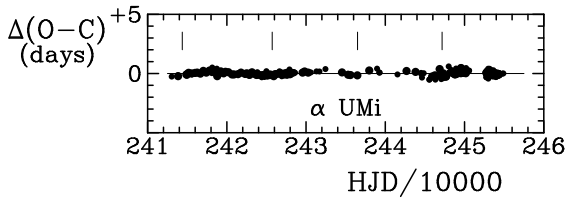}
\caption{\label{fig12} \small{The deviations of the high quality O--C data from the best-fitting parabolic trends matched to the observations. Tick marks denote epochs when the Cepheid primary should be crossing into the plane of the sky.}}
\end{figure}

Another approach can be made using the O--C data of Fig.~\ref{fig2}, since deviations from the O--C fit should also yield parameters for the binary system. The light travel time delays for a star orbiting in a binary system depend upon $a_1 \sin i$ in similar fashion to the radial velocities, although they are a quarter cycle out of phase. The largest weight O--C observations for Polaris do indeed follow such a trend, as shown in Fig.~\ref{fig12}. What is remarkable is the magnitude of the deviations, which seem to reach almost 0.15 light day, or 26 A.U., a value that seems inordinately large for the motion of the most massive star in the system about its (presumably) less massive companion.

\begin{figure}[h]
\begin{center}
\includegraphics[width=0.35\textwidth]{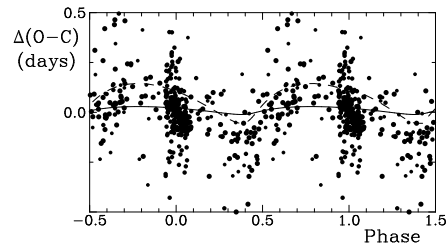}
\caption{\label{fig13} \small{Phased deviations of the O--C data from the best-fitting parabolic trends, shown relative to the orbital solution of Table~\ref{tab3} (solid line), with the size of the symbol increasing in proportion to its assigned weight. A dashed line denotes expectations for $a_1 \sin i = 5 \times$ the actual value.}}
\end{center}
\end{figure}

The last result can be seen in greater detail in Fig.~\ref{fig13}, which plots the O--C residuals as a function of orbital phase relative to expectations from the most recent orbital solution \citep{30}. It should be noted that some of the most deviant trends in Fig.~\ref{fig13} are tied to recent observations of the Cepheid from space-based platforms \citep{16,17}, so it is difficult to attribute them to ``observational error.'' A similar problem has been detected in O--C observations of the binary Cepheid RT Aur \citep{62}, although even in that case there is no satisfactory resolution of the problem by tying it to problems with the observations. Possibly the effect is simply another manifestation of the random fluctuations in pulsation period that are detected in almost all Cepheids \citep{63}, although the trends in the O--C deviations seem to be too well organized for a ``random'' process.

\section{Additional Considerations}
The numerous contradictions noted here regarding the pulsation mode of Polaris suggest the possibility of a more fundamental problem. Some of the anomalies in the radial velocity measures may be a consequence of different sets of spectral lines (ionization and excitation states) and wavelength coverage used to infer the photospheric motion, or to the inevitable problems of establishing reliable systemic velocities for a pulsating star, where shock effects are detected at some phases \citep{31}. But that does not resolve the problems arising from the O--C offsets for the photometric observations, which are of relatively high precision for recent data sets. The presence of a third star in the spectroscopic subsystem, as surmised by the results of most previous radial velocity studies \citep[e.g.,][]{33,31,32}, might resolve the various discrepancies in our current understanding of Polaris as a Cepheid, but would require continued spectroscopic and astrometric monitoring of the star for some years to come in order to provide confirmation. It is interesting to note that the ``glitch'' in the O--C observations for Polaris occurred only a few years after periastron passage for the spectroscopic subsystem in 1958, when the stars were near closest approach to one another. Presumably that would be the most likely interval of time when an interaction with another close companion would occur. But, of course, that is merely speculation. 

The reasonable possibility that some of the stars near Polaris might be physically associated with the Cepheid was noted by \citet{64}, but has not been pursued to a great extent, possibly because of the difficulties of observing stars in close proximity to the north celestial pole using ground-based facilities. The lack of detectable X-ray emission from the faint red stars C and D in the Polaris multiple system has been used by \citet{65} as evidence that they are not young, newly-arrived, main sequence dwarfs, as required by a possible association with Polaris, so are most likely field stars. There are several other brighter, potential members of the Polaris cluster lying outside the immediate field of the Cepheid, however, for which basic photometric and radial velocity observations do not exist. A program to obtain such data has been underway for a few years now, but the study is far from complete. Other low-amplitude sinusoidal Cepheids, the ``Polaris kin,'' are even less well-studied, although they could prove to be equally valuable for clarifying the nature of Polaris itself. Clearly the Polaris story is not yet over.

\end{document}